\documentstyle[aps,preprint]{revtex}
\begin{document}

\title{Polymer Adsorption on Disordered Substrate}
\author{T. Hwa and D. Cule}
\address{Physics Department\\
University of California at San Diego\\
La Jolla, CA 92093-0319}
\date{\today }
\maketitle


\begin{abstract}

We analyze the recently proposed  "pattern-matching" phase of a Gaussian
random heteropolymer
adsorbed on a disordered substrate [S. Srebnik, A.K. Chakraborty and E.I. Shakhnovich, 
Phys. Rev.  Lett. {\bf 77}, 3157 (1996)].
By mapping the problem to that of a directed homopolymer in higher-dimensional
random media, we show that the pattern-matching phase is asymptotically weakly 
unstable, and the large scale properties of the system are given by that of an
adsorbed homopolymer.

~
\end{abstract}

\newpage

In a recent letter\cite{scs}, Srebnik, Chakraborty and Shakhnovich (SCS)
presented an interesting study of the adsorption of a {\em Gaussian} random
heteropolymer (RHP) onto a disordered 2D substrate. SCS used a mean-field
analysis of a replica calculation to show that for large substrate
disorders, the polymer is adsorbed onto the substrate and forms a {\em glass}
phase which is dominated by the matching of ``patterns'' of the
RHP and the substrate. In this comment, we describe how such a pattern
matching phase might arise in principle, 
and show that it cannot exist for the  model considered by SCS.

We recall the model RHP Hamiltonian used in Ref.~\cite{scs} 
\begin{equation}
\beta {\cal H}=\int_0^Ndn\,\left\{ {\bf \dot{r}}^2+k\left[ {\bf r}(n)\right]
\theta (n)U(z(n))\right\}   \label{H}
\end{equation}
where ${\bf r}(n)=({\bf r}_{||}(n),z(n))\in \Re ^3$ and 
$\theta (n)\in \left\{ \pm 1\right\} $ denote respectively the position
and ``charge'' of the $n^{th}$ monomer,
${\bf \dot{r}}\equiv d{\bf r}/dn$, and $U(z)$ is a short-ranged
substrate potential (taken here to be of unit strength). 
The RHP is assumed to be charge-neutral on average,
with short-range correlations $\overline{\theta (n)\theta (n^{\prime })}%
=\sigma _2^2\,\delta (n-n^{\prime })$. The substrate
potential is also assumed to be short-range correlated, with $\overline{k(%
{\bf r})k({\bf r}^{\prime })}=\sigma _1^2\,\delta ^2({\bf r}_{||}-{\bf r}%
_{||}^{\prime })$. 

Consider first the case of strong disorders $(\sigma _1\cdot \sigma _2\gg 1)$ 
such that the polymer is tightly bounded to the substrate. In
this limit, the polymer conformation is controlled by the effective random
potential $W({\bf r}_{||},n)\equiv \theta (n)\,k({\bf r}_{||})$,
whose leading moments are
$\overline{W}=0$ and $\overline{W({\bf r}_{||},n)W({\bf r}_{||}^{\prime
},n^{\prime })}=\sigma _1^2\sigma _2^2\,\delta ^2({\bf r}_{||}-{\bf r}%
_{||}^{\prime })\,\delta (n-n^{\prime })$.
The explicit $n$-dependence in $W$ makes 
the problem distinct from homopolymer adsorption, and is the key ingredient
to the generation of heteropolymer effects such as pattern matching. 
Properties of the conjectured
 pattern matching phase can be obtained by viewing the
Gaussian RHP as a $(2+1)$-dimensional {\em directed polymer} in the random
potential $W({\bf r}_{||},n)$; the latter problem has been characterized 
extensively~\cite{dp}: For example,
the radius-of-gyration exponent is $\nu \approx 5/8$, and the energy cost $%
\delta E$ of confining the polymer to a region of linear size $\ell \ll
N^\nu $ crosses over from the thermal form $\delta E \sim 1/\ell^2$
to $\delta E\sim \sigma _1\sigma _2/\ell ^\omega $ for large $\ell$, with
$\omega =(2-2\nu )/\nu \approx 1.2$.

However, due to the fact that the effective potential $W$ is a product of
two {\em lower} dimensional random variables, $\theta(n)$ and $k({\bf r})$,
$W$ cannot truly be a 3-dimensional random variable. It is therefore important
to check whether the parasitic long-range correlations in $W$ (to be given
below) would modify the asymptotic behaviors, i.e., whether they are relevant
in the renormalization-group sense. The relevancy of parasitic correlations
depends on the context of the problem. We have found that such correlations
are irrelevant in a number of recent studies 
(see, e.g. Ref.~\cite{reptation}). 
Below, we show that
such correlations are instead {\em relevant} for the problem at hand.

Long-range correlations are manifested in the higher moments of $W$. 
For instance, 
the four-point correlator 
$$
\overline{W({\bf r}_{||},n)W({\bf r}_{||},n^{\prime })W({\bf r}%
_{||}^{\prime },n)W({\bf r}_{||}^{\prime },n^{\prime })} 
=\overline{k^2({\bf r}_{||})k^2({\bf r}_{||}^{\prime })}
$$
is independent of $n$ since $\theta ^2(n)=1$. This suggests the
generation of an $n$-independent potential, $V({\bf r}_{||})\sim k^2({\bf r}%
_{||})$, the existence of which can also be verified 
for a generic distribution of $\theta $ and is not
limited to $\theta \in \left\{ \pm 1\right\} $. 
The behavior of the polymer is therefore determined by the
combined effect of the potentials $V$ and $W$ which compete with each
other: $V({\bf r}_{||})$ attempts to localize the polymer to a
region of the substrate where $V$ is large and negative, while 
$W({\bf r}_{||},n)$ delocalizes the polymer (with 
$\nu > 1/2$) and encourages ``pattern
matching''. 
Quantitatively, an energy gain of the order 
$\delta V\sim \sigma _1^2/\ell^{d_{||}/2}=\sigma _1^2/\ell $ 
is obtained by localizing the polymer within a 
scale $\ell $ in a favorable region of $V$. Comparing
 $\delta V$ with the confinement cost $\delta E$ given above for the
pattern matched phase,
 one sees that 
localization effects always dominate at large scales. Optimizing the total
energy yields a confinement scale 
$\ell \sim \left( \sigma _2/\sigma _1\right) ^{1/(\omega-1)}$. 
Thus, heterogeneity is asymptotically {\em irrelevant}, 
and the interesting pattern matching phase cannot exist
for long Gaussian polymers exceeding a crossover length 
 $N_{\times }=\ell^{1/\nu }$.

So far, we discussed only the strong-adsorption limit where the polymer
lies completely on the surface. For weaker adsorptions, segments of the
polymer will escape from the surface (for both energetic and entropic 
reasons), forming a layer of loops. However, as long as the loops have
finite lengths (which is the case in the adsorbed phase by definition), 
the problem with/without loops should become equivalent upon coarse graining
beyond the loop size. We thus expect the absence of the pattern-matching 
phase throughout the adsorbed phase for the Gaussian polymer. 
Extensive numerical calculations of the
full model (\ref{H}) strongly supports the above conclusion
and will be presented elsewhere.

We are grateful to David R.~Nelson for discussions. This research is
supported by an ONR young investigator award and an A.P. Sloan research
fellowship.

%
%
%
%

\end{document}